# Nematic Colloidal Swarms Assembled and Transported on Photosensitive Surfaces

Sergi Hernàndez-Navarro, Pietro Tierno*, Jordi Ignés-Mullol, and

Francesc Sagués

*Abstract*—We demonstrate a novel method to assemble and transport swarms of colloidal particles by combining liquid crystals enabled electrophoresis and photo-sensitive surface patterning. Colloidal particles are propelled in a nematic liquid crystal via application of an alternating current electric field. Swarms of particles are assembled into a rotating mill cluster, or moved as a whole along predefined paths photo-imprinted on chemically functionalized substrates. This technique represents an alternative approach to fluid based lab-on-a-chip technologies guiding the motion of large ensembles of micrometer scale solid or liquid inclusions.

## I. Introduction

ARTIFICIAL micromachines capable of navigating in viscous fluids have great potential towards the development of novel and non invasive biomedical applications [1], the transport and delivery of bio-chemical cargos on the micro/nano scale [2], or the miniaturization of simple operations in microfluidics devices [3]. These machines can be either engineered by growing different types of materials [4], or by assembling [5] or modifying [6] colloidal units. Moreover, they are usually remotely controlled by an external field such as magnetic [7] or electric [8] ones. Colloidal particles are microscale units easily addressable by external fields, which can be used as propelling micromachines when a proper strategy is adopted in order to induce or to guide their motion in the fluid medium. In this context, most of the prototypes realized until now [9] have been based on microscopic objects suspended in water, propelled either by magnetic or electric fields. Besides recent examples with passive [10] or active [11] particles driven by a chemical reaction [12] or an external field [13], the possibility to control a large ensemble of particles along a non lithographically defined track still remains a challenging issue.

Here we show that anisometric colloidal particles immersed in a nematic liquid crystal (NLC) and subjected to an external electric field can be assembled into a rotating mill cluster upon suitable combination of chemical functionalization and light illumination. In contrast to previous approaches based on the use of water as a dispersing medium, we used liquid-crystal-enabled electrophoresis (LCEEP) [14] to move colloidal inclusions at a defined speed inside a NLC layer. Control over directionality during transport is achieved by manipulating the orientation of the NLC director with a photosensitive anchoring layer.[15] The particles can be assembled into localized swarms at arbitrary locations, moved along predefined tracks or collected into properly defined microfluidic boundaries.

## II. Experimental Methods

As colloidal units we used polystyrene particles having an anisometric (pear-like) shape (lateral dimensions $d_x$=3μm and $d_y$=4μm) and composed by two connecting spherical lobes (Magsphere Inc). These particles were dispersed in a nematic liquid crystal (NLC) MLC-7029 (Merck), characterized by a negative dielectric anisotropy $\varepsilon_a$ = -3.6 (at 1KHz). The resulting dispersion was inserted via capillary action within an experimental cell composed by two conducting glass plates separated by a polyethylene terephthalate film (Mylar, Goodfellow) having a thickness of 13 or 23 microns. The glass plates had an inner conductive face due to the presence of an indium–tin oxide (ITO) layer (sheet resistance ~100Ω, VisionTek Systems). One plate was made photosensitive by functionalizing its surface with an azosilane compound, and the photosensitivity of the azosilane film was tested by recording the UV/Vis spectra. [15] The other plate was coated with a thin layer of a polyimide compound (0626 from Nissan Chemical Industries) in order to guarantee a homeotropic surface anchoring of the liquid crystal molecules.

The experimental system is schematically shown in Fig.1. It was composed by two collimated LED sources (Thorlabs, Inc.) integrated into an optical microscope (Nikon Eclipse 50iPol). To avoid unwanted isomerization of the azosilane coating, a red filter (645nm ±50nm) was used during all measurements. The blue (455nm) and UV (365nm) light were

S. Hernàndez-Navarro, J. Ignés-Mullol and F. Sagués are with the Department of Physical Chemistry, University of Barcelona, Marti Franqués 1, 08028 Barcelona, Spain.

P. Tierno is with the Department of Structure and Constituents of Matter, University of Barcelona, Avinguda Diagonal 647, 08028 Barcelona, Spain (e-mail: ptierno@ub.edu).

P. Tierno, J. Ignés-Mullol and F. Sagués are with the Institute of Nanoscience and Nanotechnology, IN²UB, University de Barcelona, Barcelona, Spain.

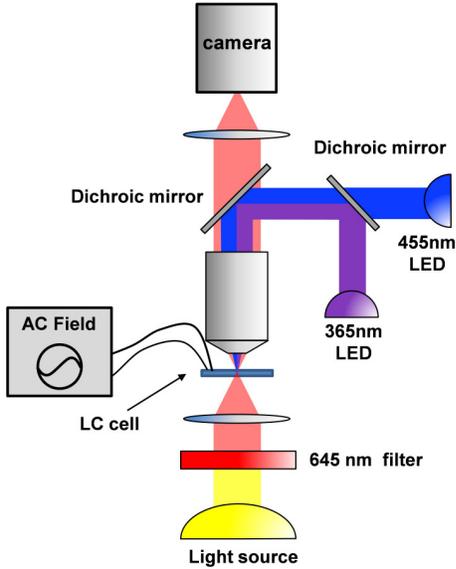

Fig. 1. Schematic illustrating the experimental system.

directed along the same path by a dichroic mirror (cutoff wavelength 405nm), while a second long-pass dichroic mirror (cutoff wavelength 505nm) was placed in front of the CMOS camera (AVT Marlin F-131B).

The microscope objective was used to focus the irradiation light onto the sample cell, reaching a power density of ~0.1W/cm$^2$ for a spot size of ~0.5 mm. Tracks were imprinted on the sample cell by moving the microscope stage while the UV light was on. The external electric field was applied by connecting the conductive side of the two plates to a voltage amplifier (TREK model PZD700) controlled by a function generator (ISO-TECH IFA 730). The range of amplitudes (frequencies) used was from 0 to 35 V peak to peak (3 to 50 Hz).

## III. PHORETIC DRIVING MECHANISM

Before discussing the collective transport of colloidal inclusions, we first describe in details the mechanism of motion of an individual particle. The dispersing medium used is a liquid crystal, i.e. an anisotropic fluid characterized by rod shaped molecules. Below a defined temperature (in our case $T \sim 95$ $^0$C), the liquid crystal is in the nematic phase, where the average molecular orientation follows a particular direction, defining a director field ***n***. The orientation of ***n*** can be controlled by proper boundary conditions at the surface of the confining cell. An external field can be used to reorient ***n*** along a defined direction. In particular, the nematic liquid crystal (NLC) used here is characterized by a negative dielectric anisotropy, and thus the nematic director orients perpendicular to the direction of the applied electric field [16,17].

When dispersed in a NLC, a colloidal particle distorts the nematic matrix creating one or more topological defects around its surface [18]. These defects can be points or lines, but as shown in Fig.2a, the anisometric particles used in this work are always characterized by two point defects, ideally located at opposite sides of the surface. These localized distortions of the LC matrix, in the form of a non-symmetric double-boojum [19], are essential for the propulsion mechanism. When the external field is applied between the two electrodes, ions start to move and redistribute around the particle surface. The non-symmetric defect structure however, creates unbalance ionic flows along the particle axis and, as a consequence, the particle starts to move along this direction [20]. Spherical particles featuring two equivalent point defects or a line defect, also called "Saturn ring" [18,19], cannot be propelled by the applied field.

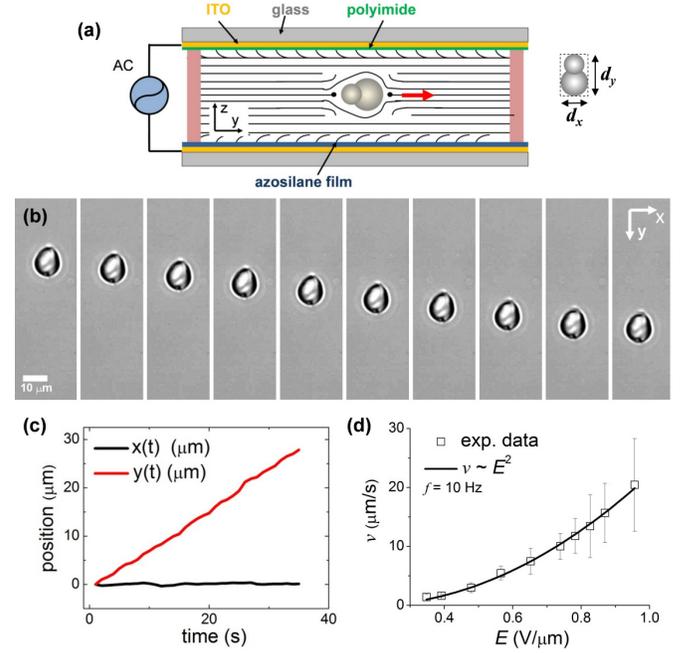

Fig.2(a) Schematic illustrating the experimental cell with one anisometric particle having lateral dimensions ($d_x,d_y$). (b) Microscope images separated by *4s* of an anisometric particle, ($d_x$=8.3 μm, $d_y$=10.2 μm), propelled by an AC field with amplitude, $E = 0.9$ V/μm and frequency, $f = 10$ Hz. (c) Corresponding position ($x,y$) versus time of the pear-shaped particle. (d) Mean velocity $v$ versus amplitude $E$ of an applied AC field with frequency $f = 10$ Hz for anisometric particles with dimensions $d_x$=3μm and $d_y$=4μm.

We test different pear-shaped particles characterized by various sizes. In most of the experiments discussed in this work, we focus on relatively small particles, having lateral dimensions $d_x$=3μm and $d_y$=4μm as illustrated in Fig.2(a). Fig. 2(b) shows a sequence of images illustrating the motion of one anisometric particle. Here we use a larger particle with dimensions $d_x$=8.3μm and $d_y$=10.2μm in order to visualize the two point defects, detected as localized dark spots on the particle surface during motion. As shown in Fig.2(c), application of the external field produces a ballistic particle transport with negligible movement in the transversal direction. Due to the large viscosity of the LC matrix, thermal fluctuations are negligible and the particle motion is completely controlled by the applied field. The moving particles always follow the nematic director, and the fact that the electric field is applied perpendicular to the sample cell completely decouples the AC electrophoretic motion from any linear (DC) electrophoretic contribution resulting from the layer. In particular, to assemble moving particles into large rotating swarms we optically manipulate with UV light the azosilane layer, forcing the latter to undergo a *trans-cis* isomerization. In absence of light, the azosilane layer is in the *trans* configuration, with the molecules perpendicular to the substrate, giving rise to a homeotropic anchoring. Upon irradiation with UV light, the azosilane adopts a *cis* configuration, bending its molecules and forcing the NLC to adopt a planar anchoring, i.e. parallel to the confining plane [21]. In order to obtain a spiral-like pattern in the NLC orientation, we first erase the central region of a UV-irradiated spot by using a smaller spot of blue light, as shown in the first schematic of Fig.3(a) and small inset of Fig.3(b). Due to this configuration, the NLC director is characterized by a homeotropic configuration inside and outside a circular corona of planar alignment. When the AC field is applied, the planar alignment of the NLC molecules inside this corona extends inwards and outwards, and the director relaxes to a bend texture resulting in a spiral configuration, as shown in the second schematic of Fig.3(a) and in Fig.3(b). During this relaxation, bend distortions are favoured since this NLC is characterized by a bend elastic constant smaller than the splay elastic constant. Once the NLC assumes this configuration, the anisometric particles moving due to LCEEP are attracted towards the center of this structure, following spiral trajectories. By increasing the density, the particles dynamically organize into a rotating cluster, as shown in Fig. 3(c). In contrast, by using irradiating the central region with a spot of UV light, we observe the accumulation of colloids into an aster-like cluster, with particles at rest [15].

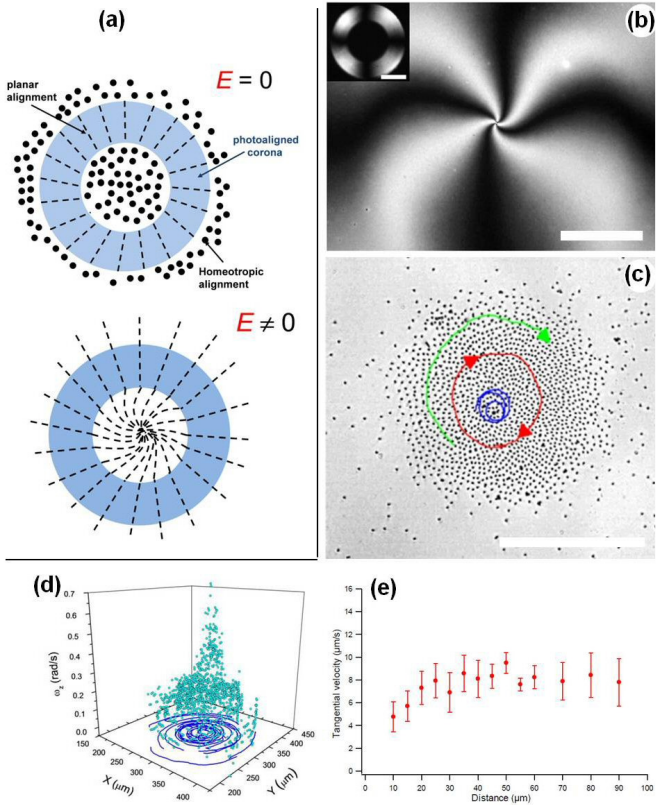

Fig. 3(a) Schematics showing the alignment of the LC molecules before and during the application of the AC field as observed from the top: points (segments) denote molecules with homeotropic (planar) alignment (b) Image from a polarized microscope with crossed polarizer and analyzer, showing the spiral pattern in the imprinted NLC texture. Inset shows a planar, photoaligned corona prior to the application of the AC field (scale bar is 500 µm). (c) Rotating mill pattern of colloidal inclusions after application of an AC field with frequency $f = 10$ Hz and amplitude $E = 0.9$ V/µm. Some particle trajectories are superimposed above the images with different colors. Scale bars in (b,c) are 200 µm. (d,e) Experimental measurements from particle tracking of the vorticity (d) and tangential velocity (e) of particles at different distances from the centre of a rotating mill. In (d) the blue lines denote the projection of the particle trajectories on the $(x,y)$ plane.

The rotating cluster composed by the moving colloidal inclusions is a mill pattern of particles which organizes in concentric trajectories, moving at constant linear velocity. As a consequence, the whole structure displays a different angular velocity on these concentric shells as one move from the center of the rotating mill towards the outer edge. We quantify attraction towards the electrodes. With this technique, it is possible to move particles up to a maximum speed of $v_{max} \sim 20$ µm/s for an applied electric field having amplitude $E = 0.9$ V/µm and frequency $f = 10$ Hz. As previously reported for liquid inclusions [20], also here the particle speed increases quadratically with the applied field $v \sim E^2$ (Fig.2(d)) for the anisometric particles, while a more complex relationship arises with the applied frequency [14]. The quadratic field dependence of the speed with the amplitude is confirmed in Fig.2(d) by fitting the experimental values (open squares) with $v = \beta(E-E_0)^2$, where we find $E_0 = 0.17$ V/µm.

## IV. ROTATING SWARMS

The possibility to control the orientation of the nematic director by modifying the surface anchoring conditions enables us to steer the particle trajectory along complex tracks designed by shining UV or blue light on the photosensitive these properties by determining the two-dimensional position (x,y) of the colloidal particles using video-microscopy and particle tracking routines. In particular we obtain the $z$ component of the vorticity $\omega_z$ by measuring the in-plane component of the particle velocity $(v_x,v_y)$ and then numerically compute the normal component of the vorticity as: $\omega_z=\partial_x v_y-\partial_y v_x$ [22]. Fig. 3(d) shows $\omega_z$ as a function of the particle position in a 3D graph (projected on the *(x,y)* plane are the particle trajectories). The vorticity reaches a maximum value of $\omega_z = 0.86$ rad/s near the center of rotation, than decreasing to $\omega_z \sim 0.15$ rad/s at a distance of ~50 μm. This result is consistent with an observed constant tangential velocity $\upsilon_\theta$ inside the cluster, as shown in Fig. 3(d). In fact both quantities are related as:

$$\omega_z = \left(\frac{v_\theta}{r} + \frac{\partial v_\theta}{\partial r}\right)\hat{e}_z$$

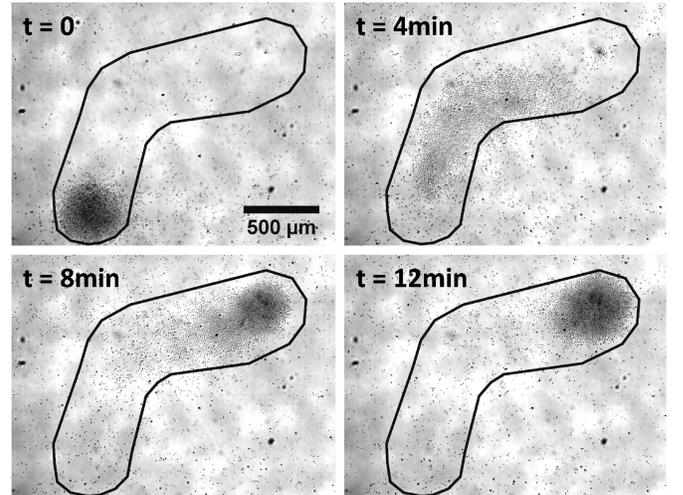

Fig. 4 Sequence of images showing a particle swarm traveling across the LC cell due to in situ reconfiguration of the LC field using UV light (365nm). The black line denotes the contour of the track, only visible between crossed polarizers. The AC field used had an amplitude of 0.74V/µm and a frequency of 10Hz.

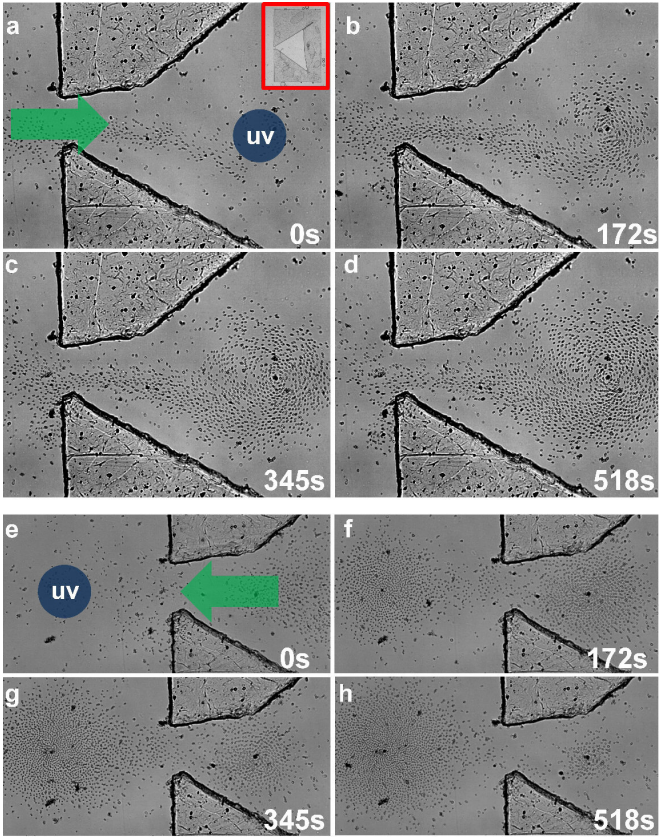

Fig.5(a-h) Sequence of images of a colloidal swarm driven inside (a-d) and outside (e-h) a triangular confinement. The shaded blue circles in panel (e) and panel (f) denote the region illuminated by UV light. Inset on top of panel (a) shows an enlargement of the entire rectangular structure with the triangular opening.

Small deviations from the linear behavior can be attributed to imperfections present in the experimental system.

## V. PHOTO-PATTERNED CIRCUITS

Rotating swarms of colloidal inclusions can be translated along arbitrary paths above the photosensitive layer by changing the location of the UV spot. This is achieved by first blocking the LCEEP mechanism, i.e. increasing the driving frequency above 50 Hz. Later the centre of attraction of the cluster of particles (.i.e. the UV spot) is translated along a predefined path, described by moving the microscope objective which imprints the prescribed trajectory. Once the frequency has been lowered again, LCEEP reactivates, and the anisometric particles start moving towards the new center of attraction following the predesigned path.

Figure 4 shows the translocation of a large swarm of particles under the application of an AC field of amplitude 0.74V/μm and frequency 10Hz, along an L-shape track. During these operation there is a minimum dismantlement of the cluster structure, and these swarms displacement can be extended on any arbitrary track on the substrate.

## VI. MICROFLUIDIC IMPLEMENTATION

Our technique to generate and transport swarms of colloidal inclusions above photo-imprinted surface tracks can be combined with the potentiality of microfluidics devices. As an example, we fabricate a rectangular structure having a size of 3×4.5 mm$^2$ and a triangular hole in it with an orifice ~100 μm wide. Fig.5(a-d) illustrates a sequence of images taken at different time intervals, where a rotating cluster formed outside the structure was subsequently driven inside the triangular compartment by rearranging the position of the attractive spot, following the protocol described in the previous section. Once attracted by the new spot, the anisometric particles pass through the orifice, and start to dynamically form a new rotating cluster after ~500 s. The particles can be stored within this microfluidic confinement for months, owing to the extremely small self-diffusion coefficient in the NLC medium, $D = 10^{-3}$μm$^2$/s (dynamic viscosity of the NLC $\eta = 0.1$ Pa s). The sequence of images in Fig.5(e-f) display the complementary operation, where the swarm of particles was reactivated by switching on the AC field, and then subsequently translated outside the microfluidic confinement.

## VII. CONCLUSION

We introduce a new technique capable of assembling and transporting colloidal inclusions in a nematic liquid crystal by properly modifying with light the anchoring conditions of one of the confining substrates. The possibility to move the imprinted pattern by translating, manually or automatically the illuminating spots allows to design any complex two-dimensional path which can be used to direct the colloidal motion. In contrast to microfluidics systems where pre-designed confining walls are needed to direct the flow of matter along microscopic channel, here our system is channel-free, and it allows writing and erasing surface tracks on the same chip. On the other hand, our approach can be integrated with lab-on-chip devices, where swarms of colloidal inclusions can be easily transported and stored into photolithographically made compartments. The use of liquid crystals in microfluidc environments is currently an emerging field of research [23-26], which will benefit from alternative approaches to manipulate, assemble or transport microscopic inclusions.

ACKNOWLEDGEMENTS

We thank Patrick Oswald for the polyimide compound and Joan Anton Farrera for experimental advices. We acknowledge financial support by MICINN (Project numbers FIS2010-21924C02, FIS2011-15948-E) and by DURSI (Project no. 2009 SGR 1055). S.H.-N. acknowledges the support from the FPU Fellowship (AP2009-0974). P.T. further acknowledges support from the ERC starting grant "DynaMO" (No. 335040) and from the "Ramon y Cajal" program (No. RYC-2011-07605).